\documentclass[prl,superscriptaddress,showpacs,longbibliography,reprint]{revtex4-2} 

\usepackage[colorlinks=true,linkcolor=blue,urlcolor=blue,citecolor=blue,pdfusetitle]{hyperref}
\usepackage{color}
\usepackage[utf8]{inputenc}
\usepackage[usenames,dvipsnames]{xcolor}
\usepackage{amsmath}
\usepackage{amssymb}
\usepackage{graphicx}
\graphicspath{{graphics/}}
\usepackage{epsfig}
\usepackage{dcolumn}
\usepackage{soul}
\usepackage{bm}
\usepackage{mathrsfs}
\usepackage{multirow}
\usepackage[all]{xy}
\usepackage{pbox}
\usepackage{lipsum}
\usepackage{verbatim}
\usepackage{braket}
\usepackage{dsfont}
\usepackage{array}
\usepackage{makecell}
\usepackage{tabularx}
\usepackage{isotope}
\usepackage{xr}
\usepackage{float}
\usepackage{mathtools}

\begin{document}
\raggedbottom

\title{Stochastic resetting induces quantum non-Markovianity}
\author{Federico Carollo}
\affiliation{Centre for Fluid and Complex Systems, Coventry University, Coventry, CV1 2TT, United Kingdom}

\author{Sascha Wald}
\affiliation{Centre for Fluid and Complex Systems, Coventry University, Coventry, CV1 2TT, United Kingdom}

\date{\today}

\begin{abstract}
Stochastic resetting describes dynamics which are reinitialized to a reference state at random times. 
These protocols are attracting significant interest:  they can stabilize nonequilibrium stationary states, generate correlations in noninteracting systems, and enable optimal search strategies. 
While a constant reset probability results in a Markovian dynamics, much less is known about non-Markovian effects in quantum stochastic resetting.
Here, we analyze memory effects in these processes---examining the evolution of quantum states and of observables---through witnesses of non-Markovianity for open quantum systems. 
We focus on discrete-time reset processes, which are of particular interest as they can be implemented on existing gate-operated quantum devices.
We show that these processes are generically described by non-divisible maps and, {\it in non-classical} scenarios where the effective reset probability can become negative, can feature revivals in the state distinguishability. 
Our results reveal non-Markovian effects in quantum stochastic resetting and show that a time-dependent reset  may be exploited to engineer enhanced stationary quantum correlations.
\end{abstract}

\maketitle
Quantum systems interacting with  an external environment can be described within the framework of open quantum systems \cite{breuer2002,gardiner2004}.
For weak system-environment couplings and fast environmental relaxation timescales, the system dynamics is Markovian, i.e.~memoryless, and  captured by quantum master equations~\cite{lindblad1976,gorini1976}. 
Non-Markovian effects stem from backflows of information in the system and play a pivotal role in quantum relaxation processes~\cite{sieberer2015,araujo2019,wald2021}. They can be incorporated through dynamical equations governed by time-dependent generators or featuring memory kernels \cite{chruscinski2010,nestman2021,ivander2024,brandner2025}. 
Several approaches have been introduced to characterize these effects, and are mostly based on  the divisibility of  dynamical maps and on revivals in the distinguishability of   quantum states~\cite{breuer2009,rivas2010,rivas2014,breuer2016} (see Ref.~\cite{chruscinski2022} for a recent review).

\begin{figure}[t]
    \centering
\includegraphics[width=\columnwidth]{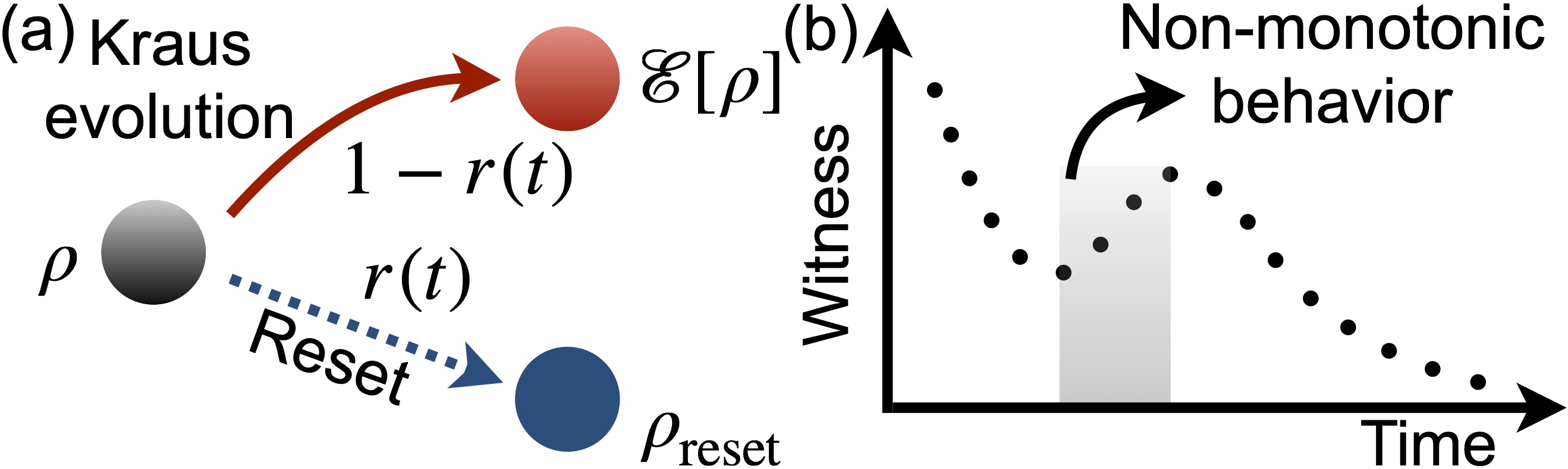}
    \caption{{\bf Quantum stochastic resetting and non-Markovianity.} (a) At any discrete time, the system state $\rho$ evolves via a Kraus map $\mathcal{E}$ or resets to a reference state $\rho_{\rm reset}$. The latter event is characterized by a reset probability $r(t)$, where $t$ is the time from the previous reset event. In genuinely quantum reset processes, $r(t)$ can assume negative values as it can happen to rates in continuous-time open quantum dynamics \cite{chruscinski2022}. (b) We explore non-Markovianity by analyzing non-monotonic behavior of certain quantities under the stochastic resetting dynamics. }
\label{fig1}
\end{figure}

Discrete-time dissipative quantum dynamics are nowadays taking on increasing importance, as they can be artificially realized on existing quantum devices~\cite{corcoles2021,deist2022,chertkov2023,graham2023,anand2024,cech2025,hothem2025}. 
In this discrete-time framework, a non-Markovian evolution can either be implemented by composing explicitly time-dependent quantum maps or through global maps which do not straightforwardly decompose into the product of time-local ones~\cite{benatti2013,garcia-perez2020,morris2022,white2025,benatti2025}.
A paradigmatic process realized in current devices is stochastic resetting~\cite{chertkov2023,wu2025} (see Refs.~\cite{evans2011,evans2020} for reset processes in classical systems).  
Quantum stochastic resetting has been predominantly studied in continuous time~\cite{rose2018,mukherjee2018,perfetto2021,wald2021a,perfetto2022,magoni2022,kulkarni2023,kulkarni2023b,sevilla2023}, with recent work extending the focus to discrete-time dynamics~\cite{yin2023,wald2025,yin2025}. 
In the latter case,  stochastic resetting unfolds as illustrated in Fig.~\ref{fig1}(a): at each time step, the system  either evolves according to an underlying Kraus map or is reset to a reference state with probability $r(t)$, depending on the time $t$ elapsed since the last reset. 
These dynamics can be easily shown to be Markovian when $r(t)$ is constant in time \cite{wald2025}. By contrast, considerably less is known about the emergence of non-Markovian effects  for explicitly time-dependent $r(t)$. This gap reflects the inherent challenges in defining and characterizing non-Markovianity in quantum dynamics \cite{addis2014,rivas2014,hou2015,breuer2016,burgarth2021,chruscinski2022,guo2022}.

In this work, we analyze discrete-time quantum reset processes and shed light on properties of their dynamical maps, in particular focussing on different notions of non-Markovianity. 
Using witnesses based on monotonicity under positive dynamics [cf.~Fig.~\ref{fig1}(b)], we characterize memory effects arising from stochastic resetting in both the Schrödinger and the Heisenberg picture \cite{settimo2025}. 
We show that quantum stochastic resetting exhibits non-Markovianity, and we derive analytical results revealing the emergence of asymptotic non-Markovian behavior in the Heisenberg picture.
Furthermore, we demonstrate that effective, negative reset probabilities $r(t)$---a scenario unique to quantum processes which parallels time-dependent dynamical generators with negative rates in specific time intervals \cite{chruscinski2022}---can yield revivals in the distinguishability of quantum states.
This  phenomenon reflects a temporary recovery of  information lost during the evolution, which is a hallmark of non-Markovianity.

Our work provides new insights on  quantum stochastic resetting and paves the way to controlling non-Markovian effects with reset processes. It further showcases a potential application of stochastic resetting with time-dependent probability $r(t)$, to the preparation of stationary states with quantum correlations exceeding what could be achieved using optimal Markovian reset strategies \cite{kulkarni2023b}. \\

\noindent \textbf{Discrete-time quantum resetting.---} 
We consider quantum systems evolving in discrete time. The system state  is described by a density matrix $\rho$, which provides the expectation value of any quantum observable $O$, as $\langle O\rangle ={\rm Tr}\left(\rho  O\right)$. 
Its dynamics is governed by the  equation 
\begin{equation}
\rho(t)=\mathcal{E}[\rho(t-1)]\, ,\quad \mbox{where} \quad \mathcal{E}[X]=\sum_{\mu}K_\mu X K_\mu^\dagger\, . 
\label{dt-me}
\end{equation}
Here, $t$ is the discrete time and the generic Kraus map $\mathcal{E}$, satisfying $\sum_\mu K_\mu^\dagger K_\mu = \mathds{1}$, implements the fundamental system update. 
Due to its structure [cf.~Eq.~\eqref{dt-me}], this map is completely positive and trace-preserving (CPTP).
Its dual $\mathcal{E}^*$ implements the evolution of quantum observables, in the Heisenberg picture, and is unital, i.e., $\mathcal{E}^*[\mathds{1}]=\sum_\mu K_\mu^\dagger K_\mu =\mathds{1}$. The  solution of Eq.~\eqref{dt-me} reads $\rho(t)=\mathcal{E}^t[\rho(0)]$, with $\rho(0)$ being the initial state. 

We assume that the above  dynamics is interspersed, at random times, by reset events. 
At any time $t$, the system has a probability to reset to a reference state $\rho_{\rm reset}$ rather than evolving via Eq.~\eqref{dt-me}, see Fig.~\ref{fig1}(a). 
This renders the dynamics stochastic 
and each realization is identified  by the times at which reset events occurred.
We focus on a process where the probability of observing a reset event at a given time solely depends on the time elapsed since the previous reset event. 
We thus introduce the probability $r(s)\in[0,1]$, for $s>0$, that a reset event occurs after $s$ time-steps from the previous one. The {\it survival probability}, i.e.~the probability that the system evolves through Eq.~\eqref{dt-me} without experiencing resetting for $t$ consecutive time-steps, is $S(t)=\prod_{s=1}^t[1-r(s)]$ with $S(0)=1$. These quantities allow us to define the probability of having a waiting time $t$  between reset events as $p(t)=r(t)S(t-1)$. 
Note that, $S(t)\to 0 $ for $t\to\infty$, if $r(t)$ does not decay too rapidly to zero, and this ensures the normalization condition $\sum_{t=1}^\infty p(t)=1$. 

In the presence of stochastic resetting, the  system dynamics, averaged over all realizations of the process, is encoded in the last {\it last renewal equation}~\cite{evans2020,magoni2022,das2022}
\begin{equation}   
\rho(t)=S(t)\mathcal{E}^t[\rho(0)]+\sum_{s=1}^t \nu(s)S(t-s)\mathcal{E}^{t-s}[\rho_{\rm reset}]\, .
\label{LRE}
\end{equation}
The first term in the above equation accounts for the reset-free dynamical realization  occurring  with probability $S(t)$.
The terms appearing in the sum in Eq.~\eqref{LRE} account, instead, for realizations in which the last reset event occurred at time $s$ and the system thereafter evolved without resetting up to time $t$. 
The quantity $\nu(s)$ is the probability that a reset event occurs exactly at time $s$. 
By exploiting trace-preservation of the dynamical map in Eq.~\eqref{LRE}, one finds the relation  
$\nu(t)=1-S(t)-\sum_{s=1}^{t-1}\nu(s)S(t-s)$, 
which can be solved iteratively starting from $\nu(1)=r(1)$. In the Supplemental Material \cite{SM}, we prove that $0\le\nu(t)\le1$ which is required for it to be a probability.

To characterize non-Markovianity in quantum stochastic resetting, we rewrite the evolution in Eq.~\eqref{LRE} as $\rho(t)=\Phi(t)[\rho(0)]$, which identifies  the dynamical map 
\begin{equation}
\Phi(t)[\cdot]=S(t)\mathcal{E}^t[\cdot]+\sum_{s=1}^t\nu(s) S(t-s) \mathcal{E}^{t-s}\circ \mathcal{P}_{\rm reset}[\cdot]\, .
    \label{dyn_map_S}
\end{equation}
Here, $\mathcal{P}_{\rm reset}[X]={\rm Tr}(X)\rho_{\rm reset}$ is the projector onto the reset state. Furthermore, we introduce the dual  map $\Phi^*(t)$, implementing the evolution of quantum observables in the Heisenberg picture as $O(t)=\Phi^*(t)[O]$. 
It reads 
\begin{equation}
\Phi^*(t)[\cdot]=S(t)\mathcal{E}^{*\, t}[\cdot]+\sum_{s=1}^t \nu(s) S(t-s) \mathcal{P}_{\rm reset}^*\circ \mathcal{E}^{*\, (t-s)}[\cdot]\, ,
    \label{dyn_map_H}
\end{equation}
where $\mathcal{P}_{\rm reset}^*[O]={\rm Tr}\left(\rho_{\rm reset }O\right) \mathds{1}$ \cite{SM}.  
The maps in Eqs.~\eqref{dyn_map_S}-\eqref{dyn_map_H} are completely positive as they are given by  convex combinations of Kraus maps  \cite{SM}. 
Moreover, $\Phi(t)$ is trace-preserving while $\Phi^*(t)$ is unital [see also discussion below Eq.~\eqref{dt-me}].
These conditions are necessary for them to describe physical quantum dynamics. \\

\noindent \textbf{Intertwining maps and divisibility.---} 
The quantum dynamical map $\Phi(t)$ can be decomposed into an evolution up to a time $s<t$,  followed by a map evolving the system from time $s$ to time $t$. Namely, we can write  
\begin{equation}
    \Phi(t)=\Lambda(t,s)\circ \Phi(s)\, , \quad \mbox{for} \quad 0\le s<t\, ,
    \label{inter_S}
\end{equation}
with $\Lambda(t,s)$ being the so-called {\it intertwining map} connecting time $s$ to time $t$.
If  $\Phi(t)$ is invertible, as it is the case, e.g., for a purely unitary dynamics, we have  $\Lambda(t,s)=\Phi(t)\circ \Phi^{-1}(s)$. 
When $\Lambda(t,s)$ is completely positive (CP), or positive (P), for all $0\le s<t$, then $\Phi(t)$ is called CP-divisible, or P-divisible respectively. 
A lack of CP- or P-divisibility has been associated with non-Markovianity~\cite{rivas2010,breuer2016,chruscinski2022}. 

Characterizing  (complete) positivity of the intertwining  maps is a prohibitively difficult task. 
In generic settings, one can typically solely rely on finding witnesses of non-divisibility \cite{benatti2013,rivas2014}. 
For the maps $\Lambda(t,s)$, acting on quantum states in the Schrödinger picture, one such witness is the trace norm $\|X\|_1:={\rm Tr}\sqrt{XX^\dagger}$. If $\Phi(t)$ is P-divisible, then $\|\Phi(t)[X]\|_1$
must be a monotonically decreasing  function of $t$ for any Hermitian operator $X=X^\dagger$ (see e.g.~Refs.~\cite{settimo2025,SM}).
A non-monotonic behavior of such a quantity then implies that $\Lambda(t,s)$, for the times in which increasing values are observed, is not positive and thus the map $\Phi$ is not P-divisible.

In the Heisenberg picture, the  intertwining maps $\Psi(t,s)$ can be analogously defined via the relation 
\begin{equation}
\Phi^*(t)=\Psi(t,s)\circ \Phi^*(s)\, , \quad \mbox{for} \quad 0\le s<t\, ,
    \label{inter_H}
\end{equation}
see the discussion in Ref.~\cite{settimo2025} for the continuous-time case. 
Since any positive unital map $\Psi$ is norm-contracting (see e.g. Ref.~\cite{SM}), i.e.~$\|\Psi[O]\|\le \|O\|$, for any operator $O$, any increase in the norm $\|\Phi^*(t)[O]\|$ for a given operator $O$ acts as witness of the non-positivity of some  intertwining map $\Psi(t,s)$ \cite{settimo2025,SM}.  
Note that, in general,  $\Psi(t,s)\neq \Lambda^*(t,s)$.
In particular, in cases in which 
$\Phi(t)$ is invertible one has $\Psi(t,s)=\Phi^*(s)\circ \Lambda^*(t,s)\circ [\Phi^{*}(s)]^{-1}$
\cite{settimo2025,SM}. As such, the two dynamical maps $\Phi(t)$ and $\Phi^*(t)$ can have very different  divisibility properties.

For completeness, we briefly consider the case of constant reset probability $r(s)=r>0$.
(This assumption provides the discrete-time equivalent of continuous-time Poissonian stochastic resetting \cite{wald2025}.) 
The  map $\Phi(t)$ reduces to the repeated application of the fundamental map $\Phi_r$. 
That is, $\Phi(t)=\Phi_r^t$ with 
\begin{equation}
    \Phi_r[\cdot]=(1-r)\mathcal{E}[\cdot]+r \mathcal{P}_{\rm reset}[\cdot]\, 
    \label{single_time_step_Markov}
\end{equation}
being completely positive. In this case, the dynamics is thus intrinsically Markovian. For time-dependent $r(t)$, the quantum stochastic resetting can already be regarded non-Markovian as it is implemented by a nonlocal map [cf.~Eq.~\eqref{dyn_map_S}], reminiscent of continuous-time equations with memory kernels \cite{chruscinski2010}. In the following, we explore properties of the intertwining maps when $r(t)$ depends on time.  \\

\noindent \textbf{Non-Markovian resetting.---}  To move away from Markovian resetting, we  consider generic probabilities $r(s)\in [0,1]$.
Exploring this setup analytically is prohibitive due to the generality of the problem.
Thus, we first present numerical results for stochastic resetting processes and then identify a class of reset processes for which we can obtain some analytic insights.

We consider several randomly generated reset processes, obtained by sampling $r(s)\in [0,0.1]$ from a uniform distribution and considering $\mathcal{E}[\cdot]=U\cdot U^\dagger$, with $U=e^{-i G}$ and $G$ is a random Hermitian operator.
We focus on the reset dynamics of random Hermitian two-qubit operators $X$, subject to the reset dynamics in Eqs.~(\ref{dyn_map_S}) and~(\ref{dyn_map_H}), with  the reset state $\rho_{\rm reset}=\ket{1}\!\bra{1}\otimes \ket{1}\!\bra{1}$. 
Here, $\ket{1}$ represents the eigenstate of $\sigma_z$  with eigenvalue $1$.
All random Hermitian operators are generated using qutip~\cite{qutip} by first generating a random matrix $A$ with a density of non-zero elements of $0.75$ and whose elements are drawn from a flat distribution, and then returning $(A+A^\dagger)/2$.

\begin{figure}[t]
    \centering
    \includegraphics[width=\columnwidth]{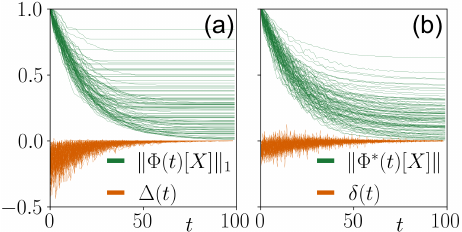}
    \caption{{{\bf P-divisibility of dynamical maps.} (a) We consider $100$ random Hermitian operators $X$ propagated, in the Schrödinger picture, via randomly generated reset processes (see main text). Their trace norm is mostly decreasing with, however, positive increments $\Delta(t)$  highlighting that $\Phi(t)$ is not P-divisible in general. (b) We generate, as in a), random operators and  processes and analyze the Heisenberg evolution. The norm $\|\Phi^*(t)[X]\|$   shows more prominent  non-monotonic behavior [see positive values of the increment $\delta(t)$], witnessing that also $\Phi^*(t)$ cannot be generally P-divisible.}}
    \label{fig:norm}
\end{figure}

To test P-divisibility for the dynamical map $\Phi(t)$, we investigate the time-dependence of $\|\Phi(t)[X]\|_1$.
In Fig.~\ref{fig:norm}(a), we show the witness $\|\Phi(t)[X]\|_1$ as well as its increment $\Delta(t)=\|\Phi(t+1)[X]\|_1-\|\Phi(t)[X]\|_1$ for a sample of 100 generated processes, defined by $r(t)$, each applied to a distinct Hermitian operator $X$.
The increment is almost always negative even though it can also assume positive values. This shows that there certainly are situations in which quantum stochastic resetting is implemented by non-P-divisible maps. 
We then consider the divisibility of the dynamical map in the Heisenberg picture $\Phi^*(t)$. 
To this end, we use the same processes and operators but we focus on the evolution of the norm $\|\Phi^*(t)[X]\|$.
In Fig.~\ref{fig:norm}(b) we consider this quantity together with its increment $\delta(t)=\|\Phi^*(t+1)[X]\|-\|\Phi^*(t)[X]\|$. 
We observe that the increment can become positive in this case as well which indicates that at certain times $t$, the intertwining maps $\Psi(t+1,t)$ is not positive. 
The signal appears to be more pronounced for the dual map in comparison to the underlying map.

To provide an example amenable to analytical calculations, we focus on the case $r(s)=r [1+\cos (s\pi )]/2$. In this case, the evolution of the state can be written as   
\begin{equation}
\rho(2t+1)=\mathcal{E}[\rho(2t)]\, \quad \mbox{and}\quad \rho(2t)=\Phi_r[\rho(2t-1)]\, ,
\label{Example_H}
\end{equation}
with $\Phi_r$ given by Eq.~\eqref{single_time_step_Markov}.
We further assume that the underlying Kraus map is purely unitary, i.e. $\mathcal{E}[\rho]=U\rho U^\dagger$, with the additional assumption that the unitary gate $U$ is also Hermitian such that $U^2=\mathds{1}$.  
We consider here a generic, pure reset state $\rho_{\rm reset}=\ket{\psi}\!\bra{\psi}$. 
The (Schrödinger picture) dynamical map $\Phi(t)$ for this process is CP-divisible, given that both $\mathcal{E}$ and $\Phi_r$ are completely positive. 
However, this does not necessarily imply that the corresponding dual map in the Heisenberg picture is divisible.
In Ref.~\cite{SM}, we show that the intertwining maps $\Psi(2t+1,2t)$ are such that 
\begin{equation}
\nonumber
    \begin{split}
        \Psi(2t+1,2t)[\ket{\psi}\!\bra{\psi}]&=U^\dagger\ket{\psi}\!\bra{\psi}U \\
        &-\frac{q(t)}{1-q(t)}\left(1-|\bra{\psi}U\ket{\psi}|^2\right) \mathds{1}\, ,
    \end{split}
\end{equation}
with $q(t)=1-(1-r)^t$. 
This relation shows that $\Phi^*(2t+1)$ is never P-divisible for $t\ge1$, given that the first term on the right hand side of the equation is a projector while the second one is, for a generic $U$, a negative multiple of the identity. The fact that the map remains non-divisible even at long times is reminiscent of the phenomenon of {\it eternal} non-Markovianity observed in continuous-time open quantum dynamics \cite{Hall14, Cres10, Megier2017, Jaga25}. \\

\begin{figure}[t]
    \centering
    \includegraphics[width =\columnwidth]{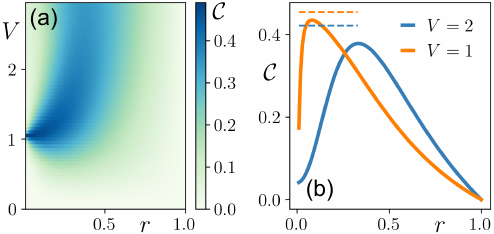}
    \caption{{\bf Engineering stationary correlations with non-Markovian reset.} (a) Concurrence $\mathcal{C}$ in the stationary state of a Markovian stochastic resetting, as a function of the interaction strength $V$ and of the reset probability $r$. Here, $\Omega=3$.  (b) For the exemplary cases $V=1$ and $V=2$, we show that a time-dependent choice $r(t)$ can recover a stationary bipartite entanglement (see dashed lines) which is higher than the one achieved by any possible Markovian stochastic resetting (solid lines).}
\label{fig3}
\end{figure} 

\noindent \textbf{State distinguishability and quantum-generalized reset.---} We now focus on a different measure of non-Markovianity for open quantum systems, which is based on the distinguishability between quantum states  \cite{breuer2009}. 
Under Markovian dynamics the distance between generic  states can only decrease so that revivals in such a quantity must be associated with a backflow of information and memory effects. 

The time-dependent distance between two quantum states $\rho_1$ and $\rho_2$ evolving through the dynamical map $\Phi$ is defined, through the trace norm, as 
\begin{equation}
\mathcal{D}(t)[\rho_1,\rho_2]:=\frac{1}{2}\left\|\Phi(t)[\rho_1-\rho_2]\right\|_1\, . 
\label{distinguishability}
\end{equation}
As mentioned above, the trace norm is non-increasing under positive maps such that Markovian dynamics only show a decreasing distance. 
Considering the dynamical map $\Phi(t)$ from Eq.~\eqref{dyn_map_S}, we see that $\mathcal{D}(t)[\rho_1,\rho_2]={S(t)}\left\|\mathcal{E}^t[\rho_1 -\rho_2]\right\|_1/2$. Given that $S(t)$ is non-increasing with $t$ [since $r(t)\ge0$] and $\mathcal{E}^t$ is Markovian, all processes considered so far classify as Markovian according to the definition based on state distinguishability.

The nonincreasing behaviour of $S(t)$ is inherently related to the fact that the reset process is essentially an external classical process and that $r(t)$ is a probability.
However, when considering the evolution in Eq.~\eqref{LRE} and the dynamical map in Eq.~\eqref{dyn_map_S} from the point of view of a quantum dynamics,  a classical interpretation of the reset process is not strictly necessary.
The only constraint for $\Phi(t)$ to be a generic quantum map is  completely positivity.
We can thus allow for certain coefficients $r(t)$ to be negative as long as the overall map $\Phi(t)$ is completely positive for all $t$.
Such a situation could result from a quantum interaction with an environment, similar to non-Markovian continuous-time setups featuring negative damping rates. 

To prove that a physical quantum dynamics $\Phi(t)$ can also be obtained with certain negative values of $r(t)$, we consider a simple class of stochastic resetting. Specifically, we consider a state ${\rho}_{\rm reset}$ which is actually the stationary state of the map $\mathcal{E}$, i.e.~$\mathcal{E}[\rho_{\rm reset}]=\rho_{\rm reset}$. Using this  and trace preservation, we find that 
\begin{equation}
\Phi(t)[\cdot]=S(t)\mathcal{E}^{t}[\cdot]+\left[1-S(t)\right]\mathcal{P}_{\rm reset}[\cdot]\, .
    \label{Phi_example}
\end{equation}
This shows that whenever $0\le S(t)\le 1$, the above map is a convex combination of two Kraus maps and is therefore completely positive.
Despite having to be smaller than $1$, nothing prevents $S(t)$ to display revivals.
In any such occasion, the quantum stochastic-resetting dynamics  shows revivals in the state distinguishability. 
For unitary $\mathcal{E}$,  we have $\mathcal{D}(t)[\rho_1,\rho_2]={S(t)}\left\|\rho_1 -\rho_2]\right\|_1/2$ which shows that $S(t)$ directly characterizes the trace distance and therefore state distinguishability. 
For generic non-unitary maps $\mathcal{E}$, instead, the behavior of $\mathcal{D}(t)$ depends on the interplay between the loss of distinguishability due to $\mathcal{E}$ and the possible gain due to revivals in $S(t)$. \\

\noindent \textbf{Engineering stationary correlations.---} Quantum reset dynamics with non-constant probability $r(t)$ have been explored previously, for instance, in relation to first hitting times \cite{yin2023} or to the existence of steady states \cite{wald2025}. Here, we take a different perspective and show that a time-dependent  $r(t)$ can allow for the engineering of stationary  quantum correlations which cannot be achieved by any Markovian stochastic resetting. 
For concreteness, we consider a two-qubit system undergoing stochastic resetting. We focus on a unitary map $\mathcal{E}[\cdot]=U\cdot U^\dagger$, with $U=e^{-i H}$ and $H = \Omega (\sigma_{\rm x} \otimes \mathds{1} + \mathds{1}\otimes \sigma_{\rm x}) + V n\otimes n$. Here,  $\sigma_{\alpha}$'s  are Pauli matrices and  $n= (\sigma_{\rm z}+\mathds{1})/2$. 
We further assume that $\rho_{\rm reset}=\ket{0}\!\bra{0}\otimes \ket{0}\!\bra{0}$ with $\sigma_{\rm z}\ket{0}=-\ket{0}$. This type of Hamiltonian is relevant in the context of Rydberg atoms \cite{saffman2010}, where also stochastic resetting to the ground state $\ket{0}$ can readily be implemented \cite{anand2024}. 

In Fig.~\ref{fig3}(a), we display bipartite entanglement between the two qubits, as quantified by the concurrence \cite{hill1997}, in the presence of a constant $r(t)=r\in[0,1]$ and $\Omega = 3$.
The plot shows that for $V \gtrsim 0.5$, there exists an optimal reset probability $r$, for which the process features a stationary state with the largest entanglement~\cite{kulkarni2023b}. 
In Fig.~\ref{fig3}(b), we  show that, by choosing specific sequences of the probability $r(t)$, it is possible to generate stationary states with a bipartite entanglement which is larger than the maximal one that can be obtained via a Markovian stochastic resetting. For $V=1$, we have used a linear ramp $r(t)= {\rm min}\{0.08,10^{-3} \cdot (t-1)\}$ while for $V=2$ we have used $r(t)=[1-\cos (t-1)]/2$. \\

\noindent \textbf{Discussion.---}  We have explored dynamical properties of discrete-time quantum dynamics subject to stochastic resetting.
Despite its fundamentally classical nature, we found that stochastic resetting can induce quantum non-Markovian effects.
We observe these effects through the divisibility properties of the corresponding quantum dynamical maps,  both in the Schrödinger and in the Heisenberg picture. 
Furthermore, we have demonstrated that quantum dynamics obtained by allowing an effective reset probability to become negative can also lead to revivals in the distinguishability of quantum states. 
In this situation the reset process loses its classical  interpretation and is close in spirit to non-Markovian open quantum dynamics with ``negative rates".
Finally, we have shown that time-dependent reset probabilities can be used to engineer strong quantum correlations that are not attainable within optimal Markovian resetting schemes. We note that our analysis directly generalizes to continuous-time dynamics, since these are  obtained by taking an infinitesimal $r(t)\propto {\rm d}t$ and a map $\mathcal{E}$ describing an infinitesimal update. 
In the future, it would be interesting  to explore properties of non-Markovian unravellings \cite{strunz2002,breuer2004} of quantum stochastic resetting in relation to performance of search algorithms or in first hitting-times problems.  \\

\textbf{Data availability.---} The data displayed in the figures is
available on Zenodo \cite{zenodo2026}.

\bibliography{refs.bib}

\newpage

\newpage
\setcounter{equation}{0}
\setcounter{figure}{0}
\setcounter{table}{0}
\makeatletter
\renewcommand{\theequation}{S\arabic{equation}}
\renewcommand{\thefigure}{S\arabic{figure}}
\makeatletter

\onecolumngrid
\newpage

\setcounter{page}{1}
\begin{center}
{\Large SUPPLEMENTAL MATERIAL}
\end{center}
\begin{center}
\vspace{0.8cm}
{\Large Stochastic resetting induces quantum non-Markovianity}
\end{center}
 \begin{center}
 Federico Carollo$^{1}$ and Sascha Wald$^{1}$
 \end{center}
 \begin{center}
 $^1${\em Centre for Fluid and Complex Systems, Coventry University, Coventry, CV1 2TT, United Kingdom}\\
 \end{center}

\section{I. Positivity of $\nu(t)$}
As stated in the main text, the quantity $\nu(t)$ provides the probability that, considering  all possible dynamical realizations, a reset event occurred exactly at the discrete time $t$.
Hence, $\nu(t)$ must be by definition positive. 
However, we have derived $\nu(t)$ recursively, exploiting trace preservation of the dynamics which gives the relation
\begin{align}
\nu(t)=1-S(t)-\sum_{s=1}^{t-1}\nu(s)S(t-s), \quad \nu(1) = r(1).
\label{rel_supp}
\end{align}
From this relation, it is not immediate to see that $\nu(t) \geq 0$. Here we prove this fact by using a strong-induction argument.
The base step is clearly satisfied by the boundary condition $\nu(1) = r(1) \geq 0$.
We then assume that $\nu(s)\geq 0$ for all $1\leq s \leq t$ and show that this implies the positivity of $\nu(t+1)$, which  concludes the proof.

First, we exploit the relation $S(t+1) = S(t)[1-r(t+1)]$ to rewrite $\nu(t+1)$ as
\begin{align}
\nu(t+1)&=1-S(t)[1-r(t+1)]-\sum_{s=1}^{t}\nu(s)[1-r(t-s+1)] S(t-s)\\
&= \left[1-S(t) - \sum_{s=1}^{t-1}\nu(s) S(t-s)\right] -\nu(t)
+r(t+1) S(t) + \sum_{s=1}^t \nu(s) S(t-s)r(t-s+1).
\end{align}
The term in the square bracket is  $\nu(t)$, such that the only remaining terms that contribute to $\nu(t+1)$ are
\begin{align}
\nu(t+1)&= r(t+1) S(t) + \sum_{s=1}^t \nu(s) S(t-s) r(t-s+1) .
\label{nu_explicit}
\end{align}
Due to our strong-induction hypothesis, the right-hand side is a sum of positive numbers which proves the claim $\nu(t+1) \geq 0$, for all $t$.
We further note that, given that $\nu(t)$ is positive, Eq.~\eqref{rel_supp} also implies that $\nu(t)\le 1$, so that this quantity indeed represents a probability.

The physical meaning of the quantity $\nu(t)$ is transparent from Eq.~\eqref{nu_explicit}. The probability of observing a reset at time $t+1$, $\nu(t+1)$, is equal to the probability of surviving without resets up to time $t$, $S(t)$, and resetting at time $t+1$ (first term in the above equation) plus the probability of having a reset at a previous time $s$ surviving from time $s$ to time $t$ without resets, and then having a reset event. The latter terms must be summed over all previous times $s\in [1,t]$ (second term in the above equation).

\section{II. Complete positivity of the dynamical maps}
In this section, we provide details on the dynamical evolution maps and discuss their complete positivity. To this end, for the sake of completeness, we first explicitly write a representation of the projector $\mathcal{P}_{\rm reset}$ as a Kraus map. 

The reset state $\rho_{\rm reset}$ is a Hermitian positive matrix and it can be diagonalized as
\begin{equation}
\rho_{\rm reset}=\sum_{k}r_k \ket{R_k}\!\bra{R_k}\, ,
\end{equation}
where $r_k\ge0$ and $\ket{R_k}$ are the eigenvectors. Considering an orthonormal basis $\{\ket{m}\}_{m}$ (this can also be the basis made by the eigenstates of $\rho_{\rm reset}$), we can write the projector as 
\begin{equation}
\mathcal{P}_{\rm reset}[\cdot]=\sum_{k,m}r_k\ket{R_k}\!\bra{m}\cdot \ket{m}\!\bra{R_k}={\rm Tr}(\cdot)\rho_{\rm reset}\, .
\end{equation}
The projector is thus  in Kraus form  since $r_k\ge0$. One can also check that the dual of this map (acting on observables) is given by 
\begin{equation}
\mathcal{P}_{\rm reset}^*[\cdot]=\sum_{k,m}r_k \ket{m}\!\bra{R_k}\cdot \ket{R_k}\!\bra{m}={\rm Tr}\left(\rho_{\rm reset}\cdot\right)\mathds{1}\, .
\end{equation}
Having shown that $\mathcal{P}_{\rm reset}$ is a Kraus map and noticing that $\nu(s)\ge0$ for positive $r(s)$, it is straightforward to see that the maps $\Phi(t)$ and $\Phi^*(t)$, given in the main text, are obtained as the convex combination of completely positive maps and are therefore completely positive themselves.

\section{III. Inequivalence of the intertwining maps}
In this section, we provide details on the definition of the intertwining maps  both in the Schrödinger and in the Heisenberg picture. 
We further provide a proof on the monotonicity of the two functions considered for witnessing non-divisibility of these maps. 

\subsection{{a.} Dynamical maps and intertwining ones}
The map given in Eq.~\eqref{dyn_map_S} is linear, trace-preserving, and completely positive as a consequence of the properties of the Kraus map $\mathcal{E}$ and of the projector $\mathcal{P}_{\rm reset}$ (see also previous section of this Supplemental Material). The  map in the Heisenberg picture ${\Phi}^*(t)$ is defined through the relation 
$$
{\rm Tr}\left(X\, \Phi(t)[\rho(0)]\right)={\rm Tr}\left(\Phi^*(t)[X]\rho(0)\right)\, ,
$$
which needs to be valid for any $X$ and $\rho(0)$. 

\subsubsection{Relation between intertwining maps}
The intertwining map $\Lambda(t,s)$ in the Schrödinger picture is defined as the map which takes the system state from time $s$, $\rho(s)=\Phi(s)[\rho(0)]$, to time $t$, $\rho(t)=\Phi(t)[\rho(0)]$, for any possible operator  $\rho(0)$. Formally, this can be defined as in Eq.~\eqref{inter_S} and one can check that if $\Phi(s)$ is invertible, then $\Lambda(t,s)=\Phi(t)\circ \Phi^{-1}(s)$. 

By definition, the intertwining map in the Heisenberg picture, $\Psi(t,s)$, is the one that maps any observable $X$, from time $s$, $X(s)=\Phi^*(s)[X]$, to time $t$, $X(t)=\Phi^*(t)[X]$. To understand the relation between $\Lambda(t,s)$ and $\Psi(t,s)$ we consider the dual of Eq.~\eqref{inter_S} to obtain 
$$
X(t)=\Phi^*(t)[X]=\Phi^*(s)\circ \Lambda^*(t,s)[X]\, .
$$
This shows that, in general, the map $\Lambda^*(t,s)$  is not the intertwining map for the Heisenberg evolution as it sits on the right of $\Phi^*(s)$ rather than on its left. The map $\Lambda^*(t,s)$ is thus equal to $\Psi(t,s)$ only if $[\Lambda^*(t,s),\Phi^*(s)]=0$. An instructive relation can be found assuming that $\Phi^*(s)$ is invertible. In this case, we can write
$$
X(t)=\Phi^*(s)\circ \Lambda^*(t,s)\circ [\Phi^*(s)]^{-1}\circ \Phi^*(s)[X]\, ,
$$
highlighting that 
\begin{equation}
X(t)=\Psi(t,s)[X(s)]\, ,\qquad \mbox{with}\qquad \Psi(t,s)=\Phi^*(s)\circ \Lambda^*(t,s)\circ [\Phi^*(s)]^{-1}\, .
\label{relation_S_H}
\end{equation}
\subsection{{b.} Witnesses of non-P-divisibility for both Schrödinger and Heisenberg pictures}

\subsubsection{Schrödinger picture: the trace norm}
The trace norm of an operator $X$, which we denote as $\|X\|_1$, is given by $\|X\|_1={\rm Tr}\sqrt{XX^\dagger}=:{\rm Tr}|X|$. 
When considering any Hermitian operator $X$,  this quantity can only decrease under the action of a positive and trace-preserving map $\Gamma$ (see e.g. Ref.~\cite{settimo2025}), i.e.
$$
\|\Gamma[X]\|_1\le \|X\|_1.
$$
This result can be used to show that if the intertwining map $\Lambda(t,s)$ is positive then one needs to have $\|\Phi(t)[X]\|_1\le \|\Phi(s)[X]\|_1$. Indeed, assuming positivity of $\Lambda(t,s)$ and using the result in the above equation, we have that 
$$
\|\Phi(t)[X]\|_1=\|\Lambda(t,s)\circ \Phi(s)[X]\|_1\le \|\Phi(s)[X]\|_1\, .
$$
As such, finding that $\|\Phi(t)[X]\|_1>\|\Phi(s)[X]\|_1$ for $t>s$ and some operator $X=X^\dagger$ implies that the map $\Lambda(t,s)$ is not positive.

\subsubsection{Heisenberg picture: the operator norm}
The operator norm is denoted as $\|X\|$ and is given by the square root of the largest eigenvalue of $X^\dagger X$. As we now show, this quantity is non-increasing under the action of a unital, positive map $\Gamma$. To this end, let us consider the operator $\|X\|\mathds{1}-X$, which is a positive operator such that
$$
\Gamma[\|X\|\mathds{1}-X]\ge0\, .
$$
Exploiting linearity and unitality, we find
$$
\|X\|\mathds{1}\ge\Gamma[X]\, ,\quad \mbox{so that}\quad \|X\|\ge \|\Gamma[X]\|\, . 
$$
Now, we can use this result to show that if the intertwining map $\Psi(t,s)$ is positive than the norm of the operator can only be non-increasing with time. Indeed, we have  ($t>s$)
$$
\|X(t)\|=\|\Phi^*(t)[X]\|=\|\Psi(t,s)[X(s)]\|\le \|X(s)\|\, ,
$$
where in the last step we assumed positivity of $\Psi(t,s)$. Finding that $\|X(t)\|>\|X(s)\|$, for $t>s$, implies that the map $\Psi(t,s)$ is not positive. 

\section{IV. a class of dynamics with non-positive intertwining maps in the Heisenberg picture}
We provide details on the stochastic resetting dynamics presented in the main text, which shows non-positive intertwining maps in the Heisenberg picture. 
In the Schrödinger picture we defined the map as follows
\begin{equation}
    \Phi(t)[\rho]=   \dots \Phi_r\circ \mathcal{E}\circ\Phi_r\circ \mathcal{E}\circ\Phi_r\circ \mathcal{E}[\rho] \, ,
\end{equation}
where for even $t$ the leftmost map on the right-hand side is $\Phi_r$, while for odd $t$ it is $\mathcal{E}$. 
We  consider a unitary map $\mathcal{E}[\rho]=U\rho U^\dagger$, where we assume that $U$ is such that $U^2=\mathds{1}$, and a pure reset state $\rho_{\rm reset}=\ket{\psi}\!\bra{\psi}$. 
Conversely, the map $\Phi_r$ includes the reset process and is given by 
\begin{equation}
    \Phi_r[\rho]=(1-r)U\rho U^\dagger+r{\rm Tr}\left(\rho\right)\ket{\psi}\!\bra{\psi} \, .
\end{equation}
It is straightforward to show that the composition of these maps is given by 
$$
\Gamma[\rho]\coloneqq\Phi_r\circ\mathcal{E}[\rho]=(1-r)\rho+r {\rm Tr}\left(\rho\right)\ket{\psi}\!\bra{\psi}\, ,
$$
and that powers of $\Gamma$ implement dynamics to all even times. The latter can be computed as 
$$
\Gamma^{t}[\rho]=[1-q(t)]\rho +q(t){\rm Tr}\left(\rho\right)\ket{\psi}\!\bra{\psi}\, , \qquad \mbox {with}\qquad q(t)=1-(1-r)^t\, .
$$
The inverse of this map can also be computed explicitly and reads 
$$
\Gamma^{-t}[\rho]=\frac{1}{1-q(t)}\rho -\frac{q(t)}{1-q(t)}{\rm Tr}\left(\rho\right)\ket{\psi}\!\bra{\psi}\, .
$$
Note that the inverse maps are not in Kraus form since the second term  is negative. Recalling Eq.~\eqref{relation_S_H} and the process considered in this section, we have $\Psi(2t+1,2t)=[\Gamma^*]^t\circ \mathcal{E}^* \circ [\Gamma^*]^{-t}$, which gives 
$$
\Psi(2t+1,2t)[X]=U^\dagger XU +\frac{q(t)}{1-q(t)}\left(\bra{\psi}U^\dagger XU\ket{\psi}-\bra{\psi}X\ket{\psi}\right)\mathds{1}\, .
$$
Taking for instance $X=\ket{\psi}\!\bra{\psi}$, we find 
$$
\Psi(2t+1,2t)[X]=U^\dagger\ket{\psi}\!\bra{\psi}U-\frac{q(t)}{1-q(t)}\left(1-|\bra{\psi}U\ket{\psi}|^2\right) \mathds{1}\, .
$$
The above operator is not positive, since the first term is a projector while the second term is a negative multiple of the identity. As such, the above map cannot be positive since $X\ge 0$ but $\Psi(2t+1,2t)[X]$ is not a positive operator. Moreover, we have that $q(t)\to 1$, for $t\to\infty$ such that the map $\Phi^*(t)$ remains asymptotically non-divisible.

\section{V. TIME-LOCAL QUANTUM STOCHASTIC RESETTING EQUATION} 
A variant of a reset process described in the main text can be defined by assuming that the reset probability ${r}(t)$ is not a function of the time elapsed from the previous reset event, but rather of the running evolution time (here also represented with $t$). This variant of the process can be  constructed through the composition of time-local dynamical maps since it does not need to carry information about the latest stochastic reset event. It thus reads 
\begin{equation}
\Phi(t)[\cdot]=\Phi_{r(t)}\circ\Phi_{r(t-1)}\circ \dots \Phi_{r(1)}[\cdot]\, ,
    \label{t_loc_reset}
\end{equation}
where $\Phi_{r(k)}$ are structurally as in Eq.~\eqref{single_time_step_Markov} but with a time-dependent $r(k)$. Here, we briefly discuss non-Markovian properties that may be expected for this process. By considering time-dependent $r(t)\ge0$, the Schrödinger dynamical map is always  CP-divisible since $\Lambda(t,s)=\Phi_{r(t)}\circ \Phi_{r(t-1)}\circ \dots \Phi_{r(s)}$. Nonetheless, the dynamical maps in the Heisenberg picture can still be not P-divisible, as clear from the fact that the example in Eq.~\eqref{Example_H} can also be realized in this setting. 
Finally, we note that the state distinguishability here would solely involve the analogous of the first term in Eq.~\eqref{LRE}, so that the analysis performed in the main text can be done also in this case. Revivals of the state distinguishability may thus only be observed in genuinely quantum reset processes where the ``probability" $r(t)$ can become negative at certain times. \\

\end{document}